\documentstyle[amssymb,aps,prl,multicol]{revtex}
\newcommand{\dd}{\text{d}}               
\newcommand{\e}{\text{e}}                
\newcommand{\tr}{\text{tr}}              
\newcommand{\gegen}{\longrightarrow}     
\newcommand{\inv}[1]{\frac{1}{#1}}       
\newcommand{\LG}{{\mathbf{G}}}           
\newcommand{\kopp}{{\text{g}}}           
\newcommand{\erww}[1]{\langle #1 \rangle}
\newcommand{\const}{\text{const}}
\newcommand{\FIL}[1]{|G_{#1}|}   
\newcommand{\breitrel}[1]{\hspace*{\tabcolsep} #1 \hspace*{\tabcolsep}}
\newcommand{\Haar}{{\text{Haar}}}
\newcommand{\lzweiskal}[2]{(#1,#2)}
\newcommand{\AbGb}{{\overline{\cal A}/\overline{\cal G}}}
\newcommand{\GR}{\Gamma}
\newcommand{\quer}{\overline}
\newcommand{\ident}{\equiv}
\newcommand{\abschnitt}{\subsection*}

\begin{document}

\title{Breakdown of the Action Method in Gauge Theories}
\author{Christian Fleischhack${}^{1,2}$\cite{Email} and Jerzy Lewandowski${}^{1,3,4}$\cite{Email2}}
\address{${}^1$Center for Gravitational Physics and Geometry,
               320 Osmond Lab, Penn State University, University Park, PA 16802 \\
         ${}^2$Max-Planck-Institut f\"ur Mathematik in den Naturwissenschaften,
               Inselstra\ss e 22--26, 04103 Leipzig, Germany \\
         ${}^3$Instytut Fizyki Teoretycznej, Uniwersytet Warszawski, 
               ul.\ Hoza 69, 00-681 Warszawa, Poland \\
         ${}^4$Max-Planck-Institut f\"ur Gravitationsphysik,
               Albert-Einstein-Institut, Am M\"uhlenberg 1, 14476 Golm, Germany 
}
\date{October 31, 2001}
\maketitle

\begin{abstract}
It is shown that the definition of physical integration measures via
``exponential of minus the action times kinematical integration measure'' 
typically contradicts properties of physical models. In particular,
theories with uncountably many non-vanishing Wilson-loop
expectation values cannot be gained this way.
The results are rigorous within the Ashtekar approach to gauge field theories.
\end{abstract}

\draft \pacs{PACS: 11.15.Tk, 02.30.Cj  
             \hspace*{\fill} 
             }
\begin{multicols}{2}
\abschnitt{Introduction}
The functional integral approach to quantum field theories
consists of two basic steps: first the determination
of a ``physical'' Euclidian integration measure $\dd\mu$
on the configuration space and second
the reconstruction of the quantum theory via an Osterwalder-Schrader
procedure. 
The latter issue has been treated rigorously in several approaches --
first by Osterwalder and Schrader \cite{OS} for scalar fields, 
recently by Ashtekar et al.\ \cite{b12} for diffeomorphism invariant theories.
However, in contrast to this, the former issue
kept a problem that has been solved completely
only for some examples.
Typically, one tried to define this integration measure $\dd\mu$ heuristically
using the action method, this means (up to a normalization factor) simply by
\[ \dd\mu := \e^{-S} \: \dd\mu_0, \]
where $S$ is the classical action of the theory under consideration and
$\dd\mu_0$ is an appropriate kinematical measure on the configuration space. 
In this letter we will discuss why just this ansatz can prevent the
rigorous description of a huge class of physical theories.
More precisely, we are going to show that in every model with uncountably
many non-vanishing Wilson-loop expectation values 
there is {\em no}\/ function $f$ {\em at all}\/ describing such a theory 
via $\dd\mu := f \: \dd\mu_0$.
The criterion above is obviously fulfilled for most of the known physical
theories, irrespectively of the dimension of the underlying space-time.
Consequently, typically the naive action method fails.

\abschnitt{Framework}
This letter is based on the Ashtekar approach \cite{a48,a30} to
gauge field theories because it is 
best-suited for solving measure-theoretical problems. 
Its basic idea goes as follows: The continuum gauge theory is known
as soon as its restrictions to all finite floating lattices are known.
This means, in particular, that the expectation values of all observables
that are sensitive only to the degrees of freedom of any fixed lattice can be 
calculated by the corresponding 
integration over these finitely many degrees of freedom.
Examples for those observables are the Wilson loop variables 
$\tr\:h_\beta$
where $\beta$ is some loop in the space or space-time and $h_\beta$
is the holonomy along that loop.

The above idea has been implemented rigorously for compact structure
groups $\LG$ as follows:
First the original configuration 
space of all smooth gauge fields (modulo gauge transforms) has been enlarged
by distributional ones \cite{a72}. 
This way the configuration space became compact and
could now be regarded as a so-called projective limit of the lattice 
configuration spaces \cite{a30}. 
These, on the other hand, consist as in ordinary 
lattice gauge theories of all possible assignments of parallel transports
to the edges of the considered
floating lattice (again modulo gauge transforms). 
Since every parallel transport is an element of $\LG$,
the Haar measure on $\LG$ yields a natural measure for the lattice theories.
Now the so-called Ashtekar-Lewandowski measure $\dd\mu_0$ \cite{a48}
is just that continuum 
measure whose restrictions to the lattice theories coincide
with these natural lattice Haar measures. It serves as a canonical
kinematical measure.
The corresponding square-integrable functions build the
Hilbert space of wave-functions.

An important class among these functions is given by the so-called
cylindrical functions \cite{a48}. These are (continuous) functions depending
only on the degrees of freedom of a finite floating lattice.
A particular example for cylindrical functions are the so-called
spin-network states \cite{e9}: Given some lattice, one labels every edge with
some representation of the structure group $\LG$ and every vertex with
some contracting 
intertwiner between the representations of the adjacent edges.
The spin network state is now defined by the 
corresponding contraction of the 
representation matrices of the parallel transports 
along the edges of the lattice.
The importance of these functions
comes from the fact that they form a complete
orthonormal system for the space of wave-functions.
Note, furthermore, that
a Wilson loop $\tr \: h_\beta$ is just a special case of a spin-network state.
Here, the underlying graph is simply one loop $\beta$ without 
self-intersections, the only edge $\beta$ is labelled with the fundamental 
representation, and the contraction at the only vertex corresponds
to taking the trace.

However, in contrast to the beautiful results in the formulation of 
quantum geometry \cite{qg9} (coupled or not with 
Yang-Mills fields \cite{e5}) within this framework, the progress to date in the 
treatment of general continuum gauge theories here has been quite limited.
Only for the two-dimensional Yang-Mills theory 
the {\em complete}\/ quantization program has been
performed explicitly \cite{b11,a6,paper1}. However, 
even there the full measure has not been defined directly 
via the action method, but using a regularization and a certain limit. 
This was necessary because no extension of the classical action 
$S = \inv4 \int F^2$ to 
distributive gauge fields is known. Probably the same problem will arise 
for more complicated models as well. 
Therefore we are going to investigate a more
fundamental problem: What kind of models at all can be studied via the action
method or might it be typical that the action method fails?

\abschnitt{Result}
Let us be given a pure gauge theory (or the pure sector of a gauge
theory), i.e.\ some integration measure $\dd\mu$ such that 
the expectation values can be computed by integrating the respective
observables over the space $\AbGb$ of all (distributive) gauge fields
modulo gauge transforms: 
\[ \erww O = \int_\AbGb O \: \dd\mu. \]
Moreover, we {\em assume}\/ that within this model there are uncountably
many non-zero spin-network expectation values. Then there is {\em no}\/
function $f$ with $\dd\mu = f \: \dd\mu_0$.

Before establishing this result, we remark that obviously
there is at least no 
$\dd\mu_0$-square-integrable function $f$. Namely, if there were
such a function $f$, it could be expanded into a generalized 
Fourier series over spin-network states $T$
(recall that these states span a complete orthonormal
basis of square-integrable functions): 
\begin{eqnarray}
f & \breitrel=  & \sum_{T} \lzweiskal T f \: T
  \breitrel\ident \sum_{T} \Bigl(\int_\AbGb \quer T f \: \dd\mu_0 \Bigr) \: T 
                            \nonumber \\
  & \breitrel=  & \sum_{T} \Bigl(\int_\AbGb \quer T \: \dd\mu \Bigr) \: T
     \breitrel=   \sum_{T} \quer{\erww T} \: \: T.
\end{eqnarray}
Here, 
$\lzweiskal{\varphi_1}{\varphi_2} := 
   \int_\AbGb \quer{\varphi_1} \varphi_2 \: \dd\mu_0$
denotes the scalar product on the Hilbert space of wave-functions.
Now, the coefficients in this series
are just the spin-network expectation values
(up to complex conjugation). Hence, by assumption,
there are uncountably many non-zero Fourier coefficients in this series, 
but this is known to be impossible for Fourier series in any Hilbert spaces.  

Let us now come to the general case und let us assume there were
some (integrable) function $f$ fulfilling $\dd\mu = f \: \dd\mu_0$.
It is well-known \cite{a48,a30} that every integrable function
can be approximated by cylindrical functions.
Let now $f_n$ be such a sequence of cylindrical functions
with $f_n \gegen f$. Since $f$ is real,
we can assume that all $f_n$ are real. (Otherwise simply take the real
part of each $f_n$.) Obviously we have 
\begin{eqnarray}
\lzweiskal{f_n}\varphi
 & \breitrel\ident & \int_\AbGb \quer f_n \varphi\:\dd\mu_0
   \breitrel=        \int_\AbGb \varphi f_n\:\dd\mu_0         \nonumber \\
\label{converg}
 & \breitrel\gegen & \int_\AbGb \varphi f\:\dd\mu_0 
   \breitrel=        \int_\AbGb \varphi \:\dd\mu 
   \breitrel\ident   \erww \varphi
\end{eqnarray}
for every continuous function $\varphi$ on $\AbGb$.
We know, by assumption, that there are uncountably many non-zero 
spin-network expectation values $\erww T$. 
But, note that we have only countably many 
approximating functions $f_n$. This means by (\ref{converg}), there must be
some $n$ such that $\lzweiskal{f_n}T$ is non-zero for uncountably many 
spin-network states $T$. But this is a contradiction, since first 
a cylindrical function that depends, say,
on the lattice $\GR$ is orthogonal to all spin-network states
that belong to a lattice different from $\GR$ and since second 
for every fixed lattice there exist only countably many spin-network states.
Consequently, there is no function $f$ with $\dd\mu = f \: \dd\mu_0$
as claimed above.

\abschnitt{Implications}
Since every Wilson loop can be interpreted as a special spin network, we get
immediately the following criterion:
If there are uncountably many non-zero Wilson-loop expectation values
in the theory under consideration, then the action method fails for that
theory, i.e.\ the definition of the integration measure via 
$\dd\mu := \inv{Z} \e^{-S} \: \dd\mu_0$ cannot yield the correct
expectation values. This is true even if we would substitute the classical
action $S$ by some other function, maybe a regularized or renormalized action.

The significance of this result comes from its wide-range applicability.
Almost all known physical gauge theories formulated in terms of loops
do have uncountably many non-vanishing Wilson-loop expectation values (WLEVs).
Already in the easiest example of a gauge theory in two dimensions, this
criterion is fulfilled. There the WLEVs of non-selfoverlapping
loops are explicitly given by
$\erww{\tr \: h_\beta} = 
        d \: \e^{-\inv2 \kopp^2 c \FIL\beta}$
with $\kopp$ being the coupling constant, 
$c$ the Casimir invariant of $\LG$, $d$ the dimension of $\LG$
and $\FIL\beta$ being the area enclosed by $\beta$
\cite{ohnenummer1,e31,d26,b11,a6,paper1}. 
Although for gauge theories in higher dimensions, such as for the pure
gauge boson sectors of QED or QCD in four
dimensions, the WLEVs are mostly not known
explicitly, the knowledge of their approximative behaviour suffices
to see that uncountably many of them do not vanish.
Hence, even there the action method fails.

But so it does typically for all theories 
describing confinement or deconfinement.
In the first case the WLEVs 
are generally expected to obey an area law \cite{Wilson,Seiler}, 
i.e.\ being more or less proportional
to $\e^{-\const \FIL\beta}$ for loops $\beta$ growing in the time-direction.
This, however, implies immediately the existence of a ``continuous'', hence
uncountable family of loops with non-zero WLEVs.
The same argumentation can be used in the case of a length law being an
indicator for deconfinement.

Finally, just the existence of some continuous (quantum) symmetry 
in the given theory should typically 
suffice for the failure of the action method.
Namely, if there is some non-vanishing WLEV,
say for the loop $\beta$, then all other expectation values for the loops
$\Phi(\beta)$ with $\Phi$ running through the symmetry group will
be non-vanishing as well. If now $\beta$ itself is not accidentally invariant
under most of the symmetries, we get again uncountably many non-vanishing 
WLEVs.

\abschnitt{Reason}
What could be the deeper reason behind that behaviour?
A striking hint comes from the observation 
that the above criterion is obviously 
{\em non}\/-applicable for gauge theories on a fixed lattice.
Since here the number of basic loops is finite, both the number
of spin networks and that of Wilson loops is infinite, but countable.
(We assume that the dimension of the structure group $\LG$ is finite.)
Therefore, the uncountability assumption above cannot be fulfilled on the level
of a finite lattice. 
And, indeed, typically one can find 
certain lattice actions $S_\GR$ such that the corresponding
lattice integration measures $\dd\mu^\GR$ can be written
as $\inv{Z_\GR} \e^{-S_\GR} \: \dd\mu^\GR_\Haar$.
However, this just means that 
the deeper reason for the breakdown of the action method
must be the continuum limit, i.e.\ the transition from a discrete space-time to 
a continuous space-time -- or, in other words, the transition from
countability to uncountability.

At the same time, this observation shows possibly the best way out 
for defining physical integration measures for gauge theories avoiding
this problem: Construct first the lattice measures using the action method
with some action adapted to the lattice, calculate then the corresponding
expectation values (if necessary, by means of certain limiting 
processes), transfer them to the continuum and reconstruct here
the full measure. In fact, this procedure has been successfully implemented,
completely, e.g., in the two-dimensional case \cite{a6,paper1,b11}. 

We only mention finally that also other attempts has been made 
using, e.g., block-spin and renormalization group techniques 
(cf.\ \cite{Balaban}), but here again
the full integration measure has not been gained in general.

\abschnitt{Remarks}
The main result of this letter can even be strengthened under 
some additional assumptions \cite{sing}. 
For that purpose, let us consider a theory
having the following three properties:
First there is a universal (bare) coupling constant, i.e.\ the 
interaction in the classical regime
between arbitrarily charged, composite matter particles
is determined completely by the interaction between the elementary particles
and the charge of the particles.
(For instance, in the electromagnetic case this simply encodes the fact that the
interaction between $n$-times charged particles equals $n^2$ the interaction
between single-charged particles.) Second there are 
some loops that are independent random variables. And third in certain
cases the WLEVs go (not too slowly) to $1$ when the considered loops shrink. 
(This assumption is met, e.g., in every model describing confinement.)

If the first two suppositions are fulfilled, then the lattice 
measures still can be received via the action method. However, if all
three conditions are given, then not only the action method fails in the
continuum, but the continuum integration measure $\dd\mu$ 
is even contained in a zero subset of the 
kinematical integration measure $\dd\mu_0$. This means, absolutely {\em none}\/
information about the physical model can be extracted from integration
over $f \: \dd\mu_0$ where $f$ is some function, maybe $f = \inv Z \e^{-S}$.

The same singularity result can be deduced
if we assume as above the existence of uncountably many non-zero WLEVs
and are provided additionally with some symmetry group that acts
ergodically on the configuration space $\AbGb$ w.r.t.\
both $\dd\mu$ and $\dd\mu_0$. \cite{Ya}
This is the case \cite{b19,b18,d46}
for the weavy states of the free Maxwell field
constructed from the polymer-like excitations of the diffeomorphism invariant
quantum theory in the canonical framework.

Finally, we note that the physical integration measure $\dd\mu$ 
is often concentrated near non-generic strata \cite{paper2+4,paper5}, i.e.\
certain singular gauge fields. \cite{sing}

\abschnitt{Conclusions}
As we have shown, the breakdown of the action method can be regarded as a
typical property of the continuum: Assuming the existence of uncountably many
non-zero Wilson loop expectation values, 
the definition of the physical interaction measure via
$\dd\mu := \inv Z \e^{-S} \: \dd\mu_0$ is {\em impossible}\/.
If one uses the action method, one can at most ``approximate''
it by lattice integration measures constructed this way. For all that it is
mostly tried to get $\dd\mu$ via the action method on the continuum level. 
Perhaps adherence to the action method is a deeper reason for the
problems with the continuum limit 
occuring permanently up to now. The desired absolute continuity
between $\dd\mu$ and $\dd\mu_0$
seems to be a deceivingly simple tool, since it hides important physical
phenomena.

But, the singularity of a measure {\em per se}\/ is completely harmless.
Namely, there is no singularity in the dual picture, i.e.\ for the expectation
values. Moreover, strictly speaking, an integration 
measure is not a physically relevant quantity; 
only expectation values are detectable. 
From the theoretical point of view it is completely sufficient 
to know that such a measure does {\em exist}\/.
Insofar, our result is just a striking hint that not the 
usage of functional integrals itself, but their definition 
is to be revised.

\abschnitt{Acknowledgements}
This work has been supported by the Reimar-L\"ust-Sti\-pen\-dium of the 
Max-Planck-Gesellschaft (CF) and in part by the
Polish CSR grant 2~P038~060~17 (JL) and the Albert-Einstein-Institut (JL).
The authors are very grateful to Abhay Ashtekar and Olaf Richter 
for their valuable comments on the draft version of this letter.


\end{multicols}



\begin{thebibliography}{99}

\bibitem[*]{Email}
e-mail: {\sf chfl@mis.mpg.de} 

\bibitem[**]{Email2}
e-mail: {\sf Jerzy.Lewandowski@fuw.edu.pl}

\bibitem{qg9}

A.~Ashtekar, {\sf e-print:\ gr-qc/9901023} and references there\-in.

\bibitem{a72}

A.~Ashtekar and C.~J. Isham, {\it Class.~Quant.~Grav.} {\bf 9},
  {1433} (1992).

\bibitem{a48}

A.~Ashtekar and J.~Lewandowski,  in {\it Knots and Quantum Gravity},
  edited by John C.~Baez (Oxford University Press, Oxford, 1994).

\bibitem{a30}

A.~Ashtekar and J.~Lewandowski, {\it J.~Math.~Phys.} {\bf 36},
  {2170} (1995).

\bibitem{d46}

A.~Ashtekar and J.~Lewandowski, {\it Class.~Quant.~Grav.} {\bf 15},
  {L117} (2001).

\bibitem{a6}

A.~Ashtekar, J.~Lewandowski, D.~Marolf, J.~Mour{\~a}o, and 
Th.~Thiemann, {\it J.~Math.~Phys.} {\bf 38}, {5453} (1997).

\bibitem{b12}

A.~Ashtekar, D.~Marolf, J.~Mour{\~a}o, and 
Th.~Thiemann, {\it Class.~Quant.~Grav.} {\bf 17}, {4919} (2000).

\bibitem{Balaban}

T. Ba\l aban, {\it Commun.~Math.~Phys.} {\bf 102}, {255} (1985);
  {\bf 119}, {243} (1988); {\bf 122}, {355} (1989); and references therein.

\bibitem{e9}

J.~C.~Baez, {\it Adv.~Math.} {\bf 117}, 253 (1996).

\bibitem{e31}

N.~E.~Brali{\'c}, {\it Phys.~Rev.~D} {\bf 22}, 3090 (1980).

\bibitem{sing}

Ch.~Fleischhack, {\it to be prepared for submission}\/.

\bibitem{paper1}

Ch.~Fleischhack, {\it J.~Math.~Phys.} {\bf 41}, {76} (2000).

\bibitem{paper2+4}

Ch.~Fleischhack, {\it Commun.~Math.~Phys.} {\bf 214}, {607} (2000).

\bibitem{paper5}

Ch.~Fleischhack, {\it J.~Geom.~Phys.\ (to appear)}.

\bibitem{ohnenummer1}

A.~A.~Migdal, {\it Zh.~Eksp.~Teor.~Fiz.} {\bf 69}, 810 (1975); 
  {\it Sov.\ Phys.\ JETP} {\bf 42}, 413. 

\bibitem{OS}

K.~Osterwalder and R.~Schrader, {\it Commun.~Math.~Phys.} {\bf 31},
  {83} (1973); {\bf 42}, {281} (1975).

\bibitem{Seiler}

E.~Seiler, {\it Gauge theories as a problem of constructive quantum field
  theory and statistical mechanics} (Springer-Verlag, Berlin, 1982).

\bibitem{e5}

Th. Thiemann, {\it Class.~Quant.~Grav.} {\bf 15}, {1281} (1998).

\bibitem{b11}

Th.~Thiemann, {\it Banach Center Publ.} {\bf 39}, {389} (1996).

\bibitem{b19}

M. Varadarajan, {\it Phys.~Rev.~D} {\bf 61}, {104001} (2000).

\bibitem{b18}

J.~M.~Velhinho, {\sf e-print:\ math-ph/0107002}.

\bibitem{Wilson}

K.~G.~Wilson, {\it Phys.~Rev.~D} {\bf 10}, {2445} (1974).

\bibitem{d26}

E.~Witten, {\it Commun.~Math.~Phys.} {\bf 141}, {153} (1991).

\bibitem{Ya}

Y.~Yamasaki, {\it Measures on infinite dimensional spaces}
(World Scientific, Philadelphia, 1985).

\end{thebibliography}
\end{document}